# Long-time persistence of hydrodynamic memory boosts microparticle transport

Sean L. Seyler
*Department of Physics, Arizona State University, Tempe, Arizona 85287, USA*

Steve Pressé*
*Department of Physics and School of Molecular Sciences,
Arizona State University, Tempe, Arizona 85287, USA*
(Dated: June 15, 2019)

In a viscous fluid, the past motion of an accelerating particle is retained as an imprint on the vorticity field, which decays slowly as $t^{-3/2}$. At low Reynolds number, the Basset-Boussinesq-Oseen (BBO) equation correctly describes nonuniform particle motion, capturing hydrodynamic memory effects associated with this slow algebraic decay. Using the BBO equation, we numerically simulate driven single-particle transport to show that memory effects persist indefinitely under rather general driving conditions. In particular, when driving forces do not vary smoothly, hydrodynamic memory substantially lowers the effective transport friction. Remarkably, this enables coasting over a spatially uneven potential that otherwise traps particles modeled with pure Stokes drag. Our results provide direct physical insight into role of particle-fluid coupling in nonequilibrium microparticle transport.

*Introduction.*—In the pioneering work of Alder and Wainwright [1], molecular dynamics simulations of hard-spheres liquid argon revealed a long-time tail in the velocity autocorrelation function, which decays as $t^{-3/2}$. Whereas standard Langevin theory predicts exponential decay over a Brownian relaxation time $\tau_B$, this algebraic decay is caused by viscous coupling between nonuniform particle motion and unsteady flow in the ambient fluid [1–3]. As a particle accelerates, the fluid it displaces also accelerates, causing a virtual mass force that augments the particle's apparent inertia. The particle also imparts momentum to the fluid as vorticity, which diffuses over a particle radius in a kinematic time $\tau_\nu$ proportional to the fluid density. A particle that translates over its own radius before the ambient fluid relaxes thus feels a force due to vorticity generated by its past motion [4]. This delayed self-interaction force—the Basset history force—underlies the slow $t^{-3/2}$ decay associated with hydrodynamic memory.

The Basset-Boussinesq-Oseen (BBO) equation [5–7] is a remarkably general description of nonuniform particle motion that captures the virtual mass and history force, both of which become significant when $\tau_\nu \simeq \tau_B$ [8]. Analytical solutions to the BBO equation, however, are generally infeasible; hence the importance of recent progress in addressing the considerable computational challenges of numerically integrating the history force [9, 10]. Indeed, at low Reynolds number, it is tempting to neglect fluid inertia *a priori*, especially since memoryless models are ubiquitous, computationally accessible, and conceptually simple. The memoryless assumption, however, tacitly implies that $\tau_\nu \ll \tau_B$; these timescales are not well separated in dense media (e.g., liquids), even when Re $\ll$ 1 [11–13]. This is indeed the case for the myriad mesoscopic phenomena at the micron scale and below, such as conformational changes in biomacromolecules [14, 15], unsteady motion of microswimmers [16, 17], colloidal flows [4, 18], and particle dispersion and sedimentation in turbulence or environmental flows [19]. Yet, without a clear picture of the physical mechanism, a frugal modeler may still reason that memory effects can be safely ignored if relevant observation timescales, such as long-time steady-state behavior, are much longer than both $\tau_\nu$ and $\tau_B$.

In this letter, we use numerical simulations to unambiguously show that neglecting the history force can lead to qualitatively incorrect particle transport under general nonequilibrium conditions. Memory effects do not merely protract the transient relaxation to steady-state but persist *indefinitely* under a significant parameter regime, implying that the BBO description may be necessary regardless observation timescales. We begin by comparing particle dynamics with and without hydrodynamic memory under time-dependent square-wave driving. The history force, which models the transient retention of latent kinetic energy in a viscous fluid bath, is found to significantly reduce the effective friction for driving frequencies below $\tau_\nu^{-1}$ by smoothing the energy transfer from an abruptly varying force to a particle. We then show that this leads to enhanced transport through a tilted potential with gaps, whereas pure Stokes dissipation tends to suppress itinerant behavior. Additionally, our results demonstrate that BBO dynamics can be efficiently modeled using an extended phase space approach wherein the memory kernel is approximated by an exponential sum [20–22].

*Problem overview.*—Consider a microsphere [23] of radius $R$, mass $m$, and density $\rho_s$ in an incompressible fluid of density $\rho$ and viscosity $\eta$, with no-slip boundary conditions. In the limit of low Reynolds number and Mach number, nonuniform motion is described by the BBO equation [5–7],

$$m_e \dot{\mathbf{v}} + \zeta_S \mathbf{v}(t) + \zeta_S \sqrt{\frac{\tau_\nu}{\pi}} \int_{t_0}^t \frac{\dot{\mathbf{v}}(\tau)}{\sqrt{t-\tau}} d\tau = \mathbf{F}(\mathbf{x}, t). \quad (1)$$

The second term—the quasisteady component—is the familiar Stokes drag with friction coefficient $\zeta_S = 6\pi\eta R$, which defines the Brownian relaxation time, $\tau_B = m_e/\zeta_S$. Unsteady flow introduces two additional forces: (1) the virtual mass force, where $m_a = \frac{2}{3}\pi R^3 \rho$ is the added mass and $m_e = m + m_a$ is the *effective mass*; (2) the Basset history force, a convolution over past acceleration. Here, the kinematic time, $\tau_\nu = R^2/\nu$, which depends on the kinematic viscosity of the fluid, $\nu = \eta/\rho$, quantifies the time for vorticity to diffuse over a distance $R$.

It is instructive to consider Eq. (1) in dimensionless form. First, we introduce the following dimensional scales, which we use in the ensuing analyses (also cf. Arminski and Weinbaum [24]):

$$t_c = \tau_B \qquad \mathcal{E}_c = k_B T \qquad v_c = \sqrt{k_B T/m_e}, \quad (2)$$

where the dimensionless time is $\tilde{t} = t/t_c$, dimensionless velocity is $\tilde{v} = v/v_c$, and so on. Accordingly, length is scaled by $L_c = v_{\text{th}} \tau_B$, mass by $M_c = m_e$, force by $F_c = k_B T/L_c$, and power by $P_c = \mathcal{E}_c/\tau_B$. By introducing $\beta = \frac{9\rho}{2\rho_s + \rho}$, one may show that the particle and fluid timescales are related by $\beta = \tau_\nu/\tau_B$ [24, 25]. Redefining $\tilde{t}$ as $t$, $\tilde{v}$ as $v$, and so on, for simplicity, one obtains the following dimensionless form:

$$\dot{\mathbf{v}} + \mathbf{v}(t) + \sqrt{\frac{\beta}{\pi}} \int_{t_0}^t \frac{\dot{\mathbf{v}}(\tau)}{\sqrt{t-\tau}} d\tau = \mathbf{F}(\mathbf{x}, t). \quad (3)$$

In the present analyses, we assume neutral buoyancy—pertinent to many processes in liquid water—so $\rho_s = \rho$, $\beta = 3$, and the history term is of order unity by Eq. (3). In this case, the history force can be neglected only if $\mathbf{F}$ varies in such a way that a particle will always be moving at nearly its terminal speed and $\dot{\mathbf{v}} \sim 0$.

There are a number of derivations of Eq. (1) [26–28], the common starting point being the *unsteady* Stokes equation [29], which must be used when $\tau_\nu/\tau_B$ is $O(1)$ [13]. Generally, $\mathbf{F}$ may consist of deterministic time-dependent and conservative components, as well as thermal fluctuations [30]; respectively, $\mathbf{F}(x, t) = \mathbf{f}(t) - \nabla U(\mathbf{x}) + \boldsymbol{\xi}(t)$. Due to the Basset memory kernel, the second fluctuation-dissipation theorem implies that $\boldsymbol{\xi}(t)$ is a colored noise process [31, 32]. In the limit of vanishing fluid inertia, $\rho_s \gg \rho$ and $\tau_\nu \ll \tau_B$, the history force vanishes, $\boldsymbol{\xi}$ becomes a white noise process, and Eq. (1) reduces to a memoryless Markovian description.

However, the full non-Markovian dynamics of Eq. (1) obfuscates the physical picture considerably. To bolster physical intuition, we focus exclusively on one-dimensional zero-temperature dynamics; finite-temperature dynamics will be treated in a follow-up study, where colored noise is naturally handled by the extended phase space approach used here [33], which is often called *Markovian embedding* [21, 22, 34]. For brevity, we refer to spherical particles obeying Eq. (1) as *BBO beads*; likewise, particles resisted only by Stokes drag (i.e., zero virtual mass and history forces, with $m_e \to m$) are called *Stokes beads*.

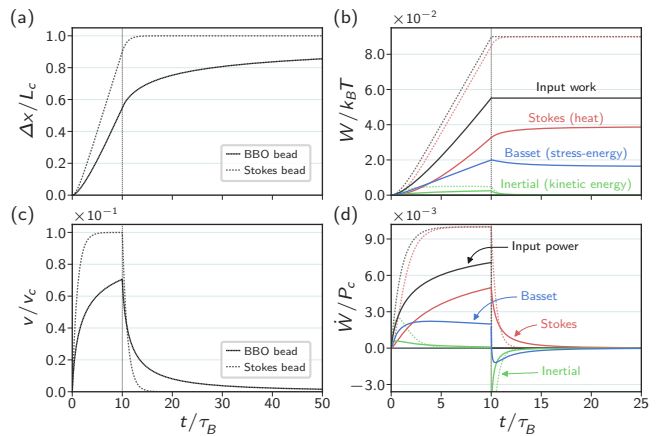

FIG. 1. While Stokes beads dissipate energy purely as heat, BBO beads partially recover energy through the Basset history term (b); for BBO beads, the Basset history term absorbs some of the input power (d), which reduces Stokes (heat) dissipation and causes net displacement (a) and instantaneous velocity (c) to vary more gradually. Pulse duration $\tau_0 = 10\tau_B$ and amplitude $f_0 = \frac{1}{10} F_c$.

*Energy flow for a constant-force pulse.*—We begin by studying BBO and Stokes beads acted upon by a finite-duration constant force. This simplest example builds intuition for the Basset history force, which can make the bead response quite complicated. Importantly, we plainly illustrate the non-dissipative nature of the history term, which can redistribute energy back to a BBO bead at later times, thereby smoothing its response to a changing external force.

Previously, Arminski and Weinbaum [24] studied the effect of pulse duration $\tau_0$ and waveform (square and triangle pulses), finding that varying $\tau_0$ altered the overall behavior more dramatically than varying the waveform did. They also showed that the maximum bead displacement $\Delta x_{\max}$ depends only on the total impulse $J_{\max} = \int F dt$ and not the dynamics, i.e., $\Delta x_{\max} = J_{\max}$; we will use this important result later. Assuming beads start and end at rest, the history term only acts to alter the details of bead motion on intermediate timescales.

Here, we consider a constant-force square pulse, $f(t) = f_0$ for $0 \leq t \leq \tau_0$ and 0 otherwise, acting on a bead starting from rest, $v(0) = 0$. We expect the long-time displacement to be proportional to $f_0 \tau_0 = \sqrt{m_e k_B T}$. Figure 1 shows results for a "long" square pulse case of duration $\tau_0 = 10\tau_B$, where $f_0 = \frac{1}{10} F_c$ and $\Delta x_{\max} = L_c$.

Due to the history force and, to a lesser extent, the virtual mass force, a BBO bead responds more gradually than a Stokes bead [Fig. 1(b)], taking tens of $\tau_B$ to reach 80% of its long-time displacement [Fig. 1(a)]. On the other hand, since the force is constant during the driving phase ($t < \tau_B$), the total input work and peak power is evidently lower for a BBO bead [Fig. 1(c,d)]. Figure 1(c) details how input work is apportioned between the instantaneous kinetic energy, total heat loss (through Stokes dissipation), and fluid stress-energy lost through vorticity diffusion (through the history term).

In particular, Stokes dissipation represents energy irreversibly lost to the bath as heat and is always positive. By contrast, the instantaneous power dissipation through the history force abruptly switches sign shortly into the relaxation phase [$t > 10\tau_B$, Fig. 1(d)], representing the transfer of stored latent energy in the fluid back to the BBO bead. This particular effect is less pronounced for short- and medium-duration pulses, though the delayed displacement at intermediate times is quite similar [35]. In any case, a BBO bead takes longer to displace than a Stokes bead but, since hydrodynamic memory improves momentum retention, less energy is required to achieve the same displacement.

*Temporal square-wave forcing.*—The single pulse results suggest that the Basset term improves the retention of momentum and energy as compared to Stokes drag alone, motivating an examination of beads driven by time-periodic forcing. However, to meaningfully compare differences between BBO and Stokes beads in transport, we should quantify the tradeoff between input energy, total displacement, and overall transport speed. To this end, we may define an effective friction coefficient $\zeta_t$ by analogy with Stokes' law,

$$\overline{F}_t = \zeta_t \overline{v}_t, \tag{4}$$

defined using the following time-averaged quantities (indicated by overbar and subscript $t$): the *effective velocity*,

$$\overline{v}_t = \frac{1}{t}\int_0^t v(\tau)\,d\tau = \frac{\Delta x_t}{t}, \tag{5}$$

and *effective driving force*,

$$\overline{F}_t = \frac{1}{\Delta x_t}\int_0^t F(x,\tau)v(\tau)d\tau = \frac{W_t}{\Delta x_t}, \tag{6}$$

where $W_t$ is net input work, $\Delta x_t = x(t) - x(0)$ is net displacement. Note that force balance in steady-state implicitly defines an effective drag via $\overline{F}_t \sim -\overline{F}_\text{drag}$. Using Eqs. (4) to (6), it is easy to compute running estimates of the effective friction.

We now consider beads driven by square waves with period $\tau = 10\tau_B$, pulse duration $\tau_0$, and amplitude $f_0$, holding fixed the impulse per cycle, $J_\text{cyc} = f_0\tau_0 = \sqrt{m_e k_B T}$. Figure 2 compares brief pulses ($\tau_0 = \frac{1}{10}\tau$ and $\frac{1}{100}\tau$) to an always-on force ($\tau_0 = \tau$). Figure 2(a) shows that BBO and Stokes beads achieve around 90% and 100% of the maximum possible displacement by $t = 250\tau_B$, consistent with the expectation that long-time displacement depends only on the total impulse, $\Delta x_\text{max} = 25 L_c$. Moreover, since fixing $J_\text{cyc}$ determines the cycle-averaged momentum input rate, we expect BBO beads to attain comparable net velocities to Stokes beads after many forcing cycles; Fig. 2(b) confirms this intuition.

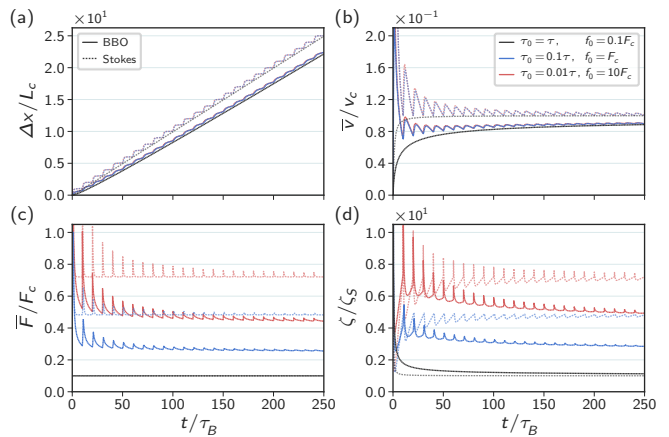

FIG. 2. For time-periodic driving with fixed impulse per cycle is fixed, BBO (solid lines) and Stokes (dotted) beads achieve comparable net displacements (a) and effective velocities (b), but it is easier to drive BBO beads (c), reducing effective friction (d). Period $\tau = 10\tau_B$; duration $\tau_0$ and force $f_0$ are varied.

For always-on forcing when the pulse duration is equal to the period, the effective driving force is identical for BBO and Stokes beads [black, Fig. 2(c)]. By contrast, increasing the abruptness of the pulses (shorter duration, larger magnitude) also increases the apparent effective driving force (i.e., increases the effective drag), but with a smaller effect on the BBO beads; however, the effective drag increases much more for a Stokes bead than a BBO bead as abruptness is increased [red/blue, Fig. 2(c)]. This behavior is expected because, as compared to a Stokes bead, a BBO bead has a lower instantaneous velocity during the forcing phase [cf. Fig. 1(c)], which reduces the total input work [cf. Fig. 1(b)]. It follows that a BBO bead's effective friction is comparable to a Stokes bead's when the driving force is always on but is lower when intermittently driven.

*Tilted spatial potential with gaps.*—We now consider beads in a tilted potential with spatial period $\lambda$ and an overall tilt $F_0$ that acts as an average driving force. We use a simple, piecewise-linear construction that alternates between downward-sloped segments (i.e., constant, positive force) and *gaps*, flat segments of length $d$ with zero force [Fig. 3, black/gray line]. This simple construction dovetails with the preceding square-wave analysis.

All beads are initialized with zero velocity at the top of a downhill segment [36]. A bead is therefore initially driven by a constant force of magnitude $F_\text{max} = F_0(\frac{\lambda}{\lambda-d})$ until reaching the first gap, by which time it has received a fixed input of energy, $W = -\Delta U_0$, where $\Delta U_0 = -F_0 \lambda$, and the bead begins coasting. It is clear a bead must coast past the first gap so as to achieve sustained transport; otherwise, it becomes localized and transport is halted. Whether a BBO or Stokes bead becomes itinerant depends on two factors dictated by its dynamics: the momentum carried by a bead upon entering the gap and the rate at which momentum is dissipated while coasting.

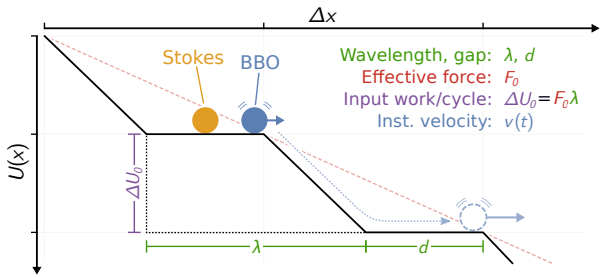

FIG. 3. Tilted periodic potential (black line) with wavelength $\lambda$, gaps of length $d$, and average slope $F_0$ (red dashes). Here, an itinerant BBO bead progresses beyond the first gap, but a Stokes bead is trapped.

Once in sustained transport, an itinerant bead receives an average energy per unit displacement (over a wavelength $\lambda$) equal to the average potential slope $F_0$, which in turn fixes the effective driving force. After an itinerant bead traverses many wavelengths ($\Delta x_t / \lambda \to \infty$ as $t \to \infty$), $\overline{F}_t \to F_0$. Recall that time-periodic driving, by contrast, essentially fixes the effective velocity but *not* the effective driving force [cf. Fig. 2(b,c)]. Accordingly, we anticipate that the effective friction in a tilted potential is predominantly determined by the effective velocity, which in turn is dictated by the dynamics.

Figure 4 depicts a representative comparison between BBO and Stokes beads for a large gap ($d = \frac{1}{2}\lambda$), small gap ($d = \frac{1}{4}\lambda$), and no gap ($d = 0$), holding wavelength ($\lambda = L_c$) and tilt ($F_0 = \frac{1}{5} F_c$) fixed. Whereas all BBO beads are able to escape, Stokes beads are trapped by the first gap, both large and small [Fig. 4(a,b)]; the final value of the effective driving force for trapped beads is set by the final displacement after coming to rest [red/blue dashed lines, Fig. 4(c)]. Furthermore, we see that the effective driving force $F_0$ approaches $\frac{1}{5} F_c$ for itinerant beads (all BBO beads and Stokes bead with no gap), in agreement with intuition [Fig. 4(c)]. Introducing gaps therefore markedly increases the effective friction of Stokes beads but has a relatively minor effect on BBO beads [Fig. 4(d)]. Varying the wavelength $\lambda$, slope $F_0$, and gap size $d$ over a wider parameter space demonstrates that this effect is general [37]. The point we wish to emphasize, however, is that it is possible construct a potential in which ordinary initial conditions lead to the localization of Stokes beads while BBO beads become itinerant.

*Discussion.*—In the tilted potential, a bead becomes itinerant when the momentum initially gained downhill is sufficient to coast past the first gap. The initial momentum input is given by the initial impulse, $J^* = F_{\max} t^*$, where the time $t^*$ to arrive at the first gap depends on the dynamics. In particular, $t^*_{\text{BBO}} \geq t^*_{\text{Stokes}}$, because a Stokes bead travels at least as fast as its BBO counterpart [Fig. 2(b)], so $J^*_{\text{BBO}} \geq J^*_{\text{Stokes}}$. Since long-time displacement only depends on $J^*$, BBO beads coast farther than Stokes beads; if BBO beads get trapped, so do Stokes beads. Hydrodynamic memory therefore always aids transport here by allowing larger gaps to be traversed.

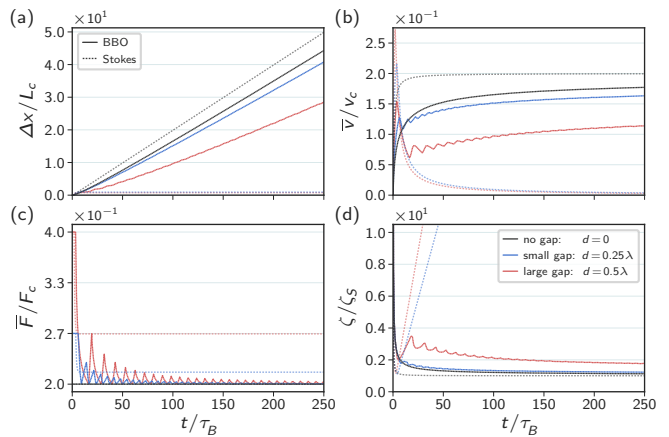

FIG. 4. Hydrodynamic memory helps BBO beads (solid lines) traverse gaps in a tilted potential, whereas Stokes beads are easily trapped (dotted). (a) Net displacement and (b) effective velocity determine, respectively, net energy and average power input. (c) The effective driving force is dictated by the slope $F_0$, thus (d) effective friction depends primarily on effective velocity. The wavelength $\lambda = L_c$ and slope $F_0 = \frac{1}{5} F_c$, so that $\Delta U_0 = \frac{1}{5} k_B T$. The initial slope is $F_{\max} = \frac{F_0}{5} (\frac{\lambda}{\lambda - d})$.

The history term retards momentum decay, smoothing out the response to fluctuations in momentum/energy input [Fig. 2(b)]. From a physical standpoint, the transient retention of momentum/energy in the fluid bath endows BBO beads with latent kinetic energy, manifesting upon rapid bead deceleration when the external force drops to zero [Fig. 1(d)]; the remaining momentum imparted to the fluid is lost at domain boundaries through vorticity diffusion. Broadly speaking, the history force—by tempering sharp instantaneous velocity gains for abruptly changing driving forces—mitigates Stokes dissipation, which grows quadratically with velocity. Moreover, for driving periods larger than roughly $\tau_\nu$ [38], the history term accumulates latent bead energy over many cycles, causing the cycle-averaged work [Fig. 2(c)] and effective friction [Fig. 2(d)] to decline over hundreds of Brownian relaxation times. This accumulation is also evident for itinerant BBO beads in the tilted potential [red line, Fig. 4(d)].

Our findings provide a foothold for understanding the compressible flow case [18, 39, 40], as well as full non-Markovian dynamics at finite temperature where hydro-dynamic memory drives long-time persistence of thermal fluctuations [41]. In view of Ref. [42], where it was shown that even tagged fluid particles closely follow BBO dynamics, the present results are germane to temperature and pressure control algorithms in molecular dynamics simulations, especially as extended phase space approaches show promise in improving the kinetic consistency of thermostats [43, 44]. Indeed, such approaches can be brought to bear on the question of stochastic transport efficiency of linear molecular motors like kinesin [45–47], with the possibility to incorporate viscoelastic fluid properties [22, 48**?**, 49].

The authors are indebted to Julian Lee, Kyle Seyler, Charles Seyler, Zeliha Kilic, David Ferry, and Ian Welland for fruitful discussions and critical feedback on the manuscript. This work was supported by ARO grant W911NF-17-1-0162 on "Multi-Dimensional and Dissipative Dynamical Systems: Maximum Entropy as a Principle for Modeling Dynamics and Emergent Phenomena in Complex Systems."

---

# Supplemental Material:
# Long-time persistence of hydrodynamic memory boosts microparticle transport

Sean L. Seyler

*Department of Physics, Arizona State University, Tempe, Arizona 85287, USA*

Steve Pressé*

*Department of Physics and School of Molecular Sciences,
Arizona State University, Tempe, Arizona 85287, USA*
(Dated: June 15, 2019)

## THE BBO EQUATION

The Basset-Boussinesq-Oseen (BBO) equation [3, 6, 17] is a Lagrangian description [4] of the velocity of a rigid sphere of radius $R$, $m$, and density $\rho_s$, moving nonuniformly through an incompressible Newtonian fluid of density $\rho$ and viscosity $\eta$. The limit of zero Reynolds and Mach number is assumed, along with no-slip boundary conditions. Expressing the velocity $\mathbf{v}$ of the sphere relative to the background fluid in terms of the relevant physical parameters, the BBO equation takes the following form:

$$(\tfrac{4}{3}\pi R^3 \rho_s)\frac{d\mathbf{v}}{dt} + (\tfrac{2}{3}\pi R^3 \rho)\frac{d\mathbf{v}}{dt} + 6\pi\eta R \mathbf{v}(t) + 6R^2\sqrt{\pi\eta\rho}\int_{t_0}^{t}\frac{d\tau}{\sqrt{t-\tau}}\frac{d\mathbf{v}}{d\tau} = \mathbf{F}(\mathbf{x}, t). \qquad (1)$$

The total external force may be decomposed into conservative, time-dependent, and fluctuating thermal force components: $\mathbf{F}(x,t) = \mathbf{f}(t) - \boldsymbol{\nabla} U(\mathbf{x}) + \boldsymbol{\xi}(t)$. In the present analyses, the thermal noise $\boldsymbol{\xi}$ is neglected in the zero-temperature limit. The physical parameters in Eq. (1) are summarized in Table I. There are two important timescales implied in Eq. (1), one corresponding to the bead and the other to the fluid (see Table I). The particle relaxation time of interest is $\tau_B$, the Brownian relaxation time. $\tau_B$ characterizes the rate of momentum damping due to Stokes drag, which gives rise to exponential decay of velocity autocorrelation function. The competing fluid timescale is the kinematic time $\tau_\nu$, which corresponds to the timescale of vorticity diffusion $\tau_\nu$ and characterizes the time to transport (shear) momentum in the fluid over a particle radius.

## NUMERICAL METHOD

The BBO equation [Eq. (1)] is challenging to solve numerically due to complications arising from the integral and, in the finite-temperature case, time-correlations in thermal fluctuations that give rise to colored noise [14]:

1. The memory kernel $t^{-1/2}$ is singular at $t = t'$, which requires care when performing direct numerical integration of the convolution term.

2. Computing the convolution integral necessitates storage of a relatively large portion of the past velocity trajectory.

3. The colored noise necessitates the generation of correlated pseudo-random number sequences with the correct statistics.

Be that as it may, there is a powerful approach known as a *Markovian embedding*, which can be used to transform a history-dependent integrodifferential equation into a numerically tractable system of equations that are *local* in time. To wit, it has been known for some time that any non-Markovian process can be represented as a projection of a higher-dimensional Markov process [7, 15, 21]. Although we are concerned with the zero-temperature case in the present study, the Markovian embedding technique naturally address the above three challenges, including the generating of the correct colored noise.

We leverage Markovian embedding to obtain an approximation to the non-Markovian GLE represented by Eq. (1) in terms of a finite set of auxiliary variables. There are a number of advantages to such an approach, as it: (1) avoids the computational (i.e. memory intensive) expense and complication of direct computation of the Basset history force integral; (2) generates, on-the-fly, the statistically correct colored noise satisfying the fluctuation-dissipation theorem [12, 16]; (3) allows the resulting (embedded) dynamical system to be propagated forward in time using standard, robust Langevin integrators, controlled to an arbitrary degree of precision; and (4) can achieve acceptable levels of statistical precision small using relatively few auxiliary variables (i.e. small embedding dimension).

Table I. Important physical parameters, characteristic scales, and units.

| name | expression | units | description |
|---|---|---|---|
| bead radius | $R$ | $L$ | — |
| bead density | $\rho_s$ | $M/L^3$ | — |
| fluid density | $\rho$ | $M/L^3$ | — |
| dynamic viscosity | $\eta$ | $M/LT$ | — |
| kinematic viscosity | $\nu = \eta/\rho$ | $L^2/T$ | — |
| bead mass | $m = \frac{4}{3}\pi R^3 \rho_s$ | $M$ | — |
| added mass | $m_a = \frac{2}{3}\pi R^3 \rho$ | $M$ | — |
| effective mass | $m_e = m + m_a$ | $M$ | — |
| Stokes friction coefficient | $\zeta = 6\pi\eta R$ | $M/T$ | — |
| Brownian collision frequency | $\gamma_B = \zeta/m_e$ | $1/T$ | — |
| Brownian/Stokes relaxation time | $\tau_B = m_e/\zeta$ | $T$ | bead momentum decay time via Stokes damping |
| kinematic time | $\tau_\nu = R^2/\nu$ | $T$ | diffusion time of fluid vorticity |
| characteristic time | $t_c = \tau_B$ | $T$ | Brownian/Stokes relaxation/damping time |
| characteristic energy | $\mathcal{E}_c = k_B T$ | $ML^2/T^2$ | thermal energy |
| characteristic mass | $M_c = m_e$ | $M$ | effective mass |
| characteristic velocity | $v_c = v_{\text{th}} = \sqrt{k_B T/m_e}$ | $L/T$ | thermal speed |
| characteristic length | $L_c = v_{\text{th}}\tau_B = \tau_B\sqrt{k_B T/m_e}$ | $L$ | — |
| characteristic force | $F_c = k_B T/L_c$ | $ML/T^2$ | — |
| characteristic power | $P_c = k_B T/\tau_B$ | $ML^2/T$ | — |
| characteristic momentum/impulse | $J_c = \sqrt{m_e k_B T}$ | $ML/T$ | — |

**Markovian embedding of BBO in the zero-temperature limit**

We follow the approach taken by Goychuk [9], Siegle et al. [19] and, more generally, Baczewski and Bond [2], wherein the algebraic form of the memory kernel is approximated as a weighted sum of exponential terms. This is known as a Prony series approximation in constitutive modeling contexts [8, 13]). This embedding leads to an *extended phase space* representation in terms of the usual position $x$ and velocity $v$, as well as a finite set of $N$ auxiliary variables $s_i$, where $i = 1, ..., N$, each obeying a stochastic differential equation (SDE) with *white* Gaussian noise. Note that the numerical system studied here will be somewhat simplified, as we can neglect the noise terms in the zero-temperature limit.

Using the fractal scaling employed by Siegle et al. [19], we obtain an approximation to the full memory kernel (which includes both the Stokes drag—which were dropped, in an *ad hoc* fashion, in the equation of motion studied by Siegle et al. [19]—as well as the history term),

$$\zeta_{\text{BB}}(t) \approx \zeta\delta(t) + \frac{C(b)}{2\sqrt{\pi}} \sum_{k=1}^{N} \sqrt{\frac{\nu_0}{b^k}} \left[ 2\delta(t) - \frac{\nu_0}{b^k} \exp\left(-\frac{\nu_0}{b^k}t\right) \right], \qquad (2)$$

$$\approx \zeta\delta(t) + \frac{C(b)}{2\sqrt{\pi}} \sum_{k=1}^{N} \sqrt{\nu_k} \left[ 2\delta(t) - \nu_k \exp\left(-\nu_k t\right) \right], \qquad (3)$$

where $\nu_k = \nu_0/b^k$, $b$ is a scaling parameter, $\nu_0$ is a high-frequency cutoff, and $C(b)$ is a fitting coefficient. The delta function in the first term comes directly from Stokes' law; the other delta function (in the sum) comes from the positive singularity in the memory kernel (when $t_0 \to 0^-$) and is necessary to ensure the integral of the kernel in the long-time limit vanishes [18, 19]. [1]

The slowly decaying power law, $t^{-3/2}$, can be approximated over $N \log_{10} b - 2$ decades through a suitable choice of auxiliary dimensions, $N$, and temporal spacing, $b$, of the exponential terms in Eq. (2) [10, 19]. The primary restriction on the high-frequency cutoff is that the timestep be small enough to achieve numerical stability when an explicit integration method is used, i.e., $\Delta t \lesssim \nu_0^{-1}$. In principle, decreasing $b$ and increasing $N$ allows the true memory kernel to be approximated to an arbitrary degree of precision. In most cases, however, one is likely to be interested in efficiency, whereby the memory kernel may be approximated by a small number of auxiliary variables that satisfactorily capture the relevant physical range of timescales.

The embedding specified via Eq. (2) transforms the linear, integro-differential BBO equation, Eq. (1), into a coupled system of $D = N + 2$ linear equations with $N$ auxiliary variables of the exponential relaxation type. The auxiliary variables comprise an $N$-dimensional extended phase space that approximates both the convolution integral representing the history force:

$$\dot{x} = v(t) \tag{4}$$

$$\dot{v} = \frac{f_{\text{ext}}(x,t)}{m_e} - \sum_{k=1}^{N} s_k(t) - \gamma_0 v(t) - \gamma_B v(t)$$

$$\dot{s}_k = -\nu_k \gamma_k v(t) - \nu_k s_k(t),$$

where $N$ is the number of auxiliary variables used to approximate the power law in Eq. (1), $k = 1, \ldots, N$, and $D = N + 2$ is the dimensionality of the embedded dynamics. For notational brevity, we have used $\gamma_B = \zeta/m_e$, which is the Brownian collision frequency, and $\gamma_0 = \sum_k \gamma_k$.

**Euler-Maruyama integration**

The equations of motion were solved using forward-Euler integration, which was sufficient for the purposes of the present study. The BBO code is implemented in the Cython language [5], which is available open source (https:\bitbucket.org/sseyler/glsimulator/src/bbo/) and will be refined in the near future to enable more general use.

The forward-Euler discretization of Eq. (4) is

$$x^{n+1} = x^n + \Delta t v^n, \tag{5}$$

$$v^{n+1} = (1 - \gamma \Delta t - \gamma_0 \Delta t) v^n + \Delta t \left( \frac{F^n}{m_e} - \sum_{k=1}^{N} s_k^n \right), \tag{6}$$

$$s_k^{n+1} = (1 - \nu_k \Delta t) s_k^n - \nu_k \gamma_k \Delta t v^n, \tag{7}$$

where $\Delta t$ is the timestep, and a superscript $n$ denotes evaluation at the $n^{\text{th}}$ timestep, and so on. We have implemented more sophisticated integration schemes, including a stochastic symplectic scheme that might be suitable for more sophisticated numerical experiments [2, 20].

A dimensionless timestep of $\Delta t = 10^{-5}$ was used and found to be more than adequate in terms of accuracy. We used $N = 12$ auxiliary variables for the single-pulse simulations [Fig. 1 in the main text], and up to $N = 21$ for longer simulations with time-periodic forcing or simulations in the tilted potential [Fig. 2–4 in the main text]. A high-frequency cutoff $\nu_0 = 10^4$ provided sufficient accuracy in all simulations. Following Siegle et al. [19], we set the scaling parameter $b = 5$ and fitting coefficient $C(b) = 1.78167$, which were sufficient to generate essentially indistinguishable trajectories from the analytical solutions (in the single-pulse case).

Finally, consider the following substitutions in the equation for velocity,

$$\theta = 1 - \gamma \Delta t - \gamma_0 \Delta t, \tag{8}$$

and, for the auxiliary equations,

$$\theta_k = 1 - \nu_k \Delta t. \tag{9}$$

---

[1] When Eq. (2) is integrated over long times ($t \to \infty$), the latter delta-function contribution must exactly cancel the total contribution from the (sum of the) exponentials [18]. Cancellation of the history term in the long-time limit is physically necessary to recover the correct (long-time) diffusive behavior—i.e., the Stokes-Einstein relation, which only depends on the Stokes friction $\zeta$ and not the kinematic time $\tau_\nu$. Furthermore, the independence of the long-time displacement of a bead (starting and ending at rest) from the dynamics—as discussed in the main text—is also a necessary consequence of the vanishing of the history term.

These substitutions, which are helpful for implementing the algorithm, enable the conversion of Eqs. (5) to (7) to the final simple form:

$$x^{n+1} = x^n + \Delta t v^n \tag{10}$$

$$v^{n+1} = \theta v^n + \Delta t \left( \frac{F^n}{m_e} - \sum_{k=1}^{N} s_k^n \right) \tag{11}$$

$$s_k^{n+1} = \theta_k s_k^n - (1-\theta_k)\gamma_k v^n. \tag{12}$$

## DECREASING THE DURATION OF A SINGLE SQUARE PULSE

In the main text, Fig. 1 shows the results for a single pulse where the pulse duration was an order of magnitude longer than the Brownian relaxation time, i.e., $10\tau_B$. This result corresponds to the long-pulse results presented in Fig. 3 of Arminski and Weinbaum [1]. When the pulse is long, the sign-change of the Basset history force is rather abrupt once the external force is removed and the bead enters the relaxation phase. We also show results for an abrupt pulse with a short duration of $\frac{1}{10}\tau_B$ [Fig. 1], which also corresponds to the short-pulse result presented in Fig. 2 of Arminski and Weinbaum [1]. For a very abrupt pulse whose duration is an order of magnitude shorter than $\tau_B$, the history force clearly dominates the Stokes drag. Thus, the energy imparted by short-timescale forces on a BBO bead will be dissipated primarily through the action of vorticity diffusion, which carries the energy away from the bead and toward the fluid boundaries [Fig. 1(b,d)].

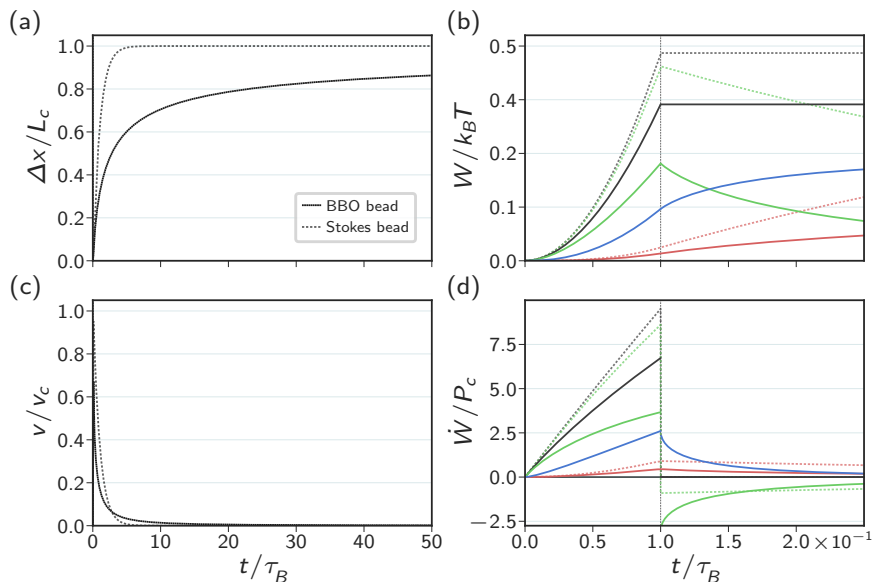

FIG. 1. Analytical solutions for BBO (solid lines) and Stokes (light dashed lines) beads for a single abrupt pulse. (a) The net displacement $\Delta x(t)$ and (c) instantaneous velocity $v(t)$ both differ in the driving and relaxation phases. (b) The net work due to external driving (black) balances the kinetic energy (green), dissipated heat (Stokes term, orange) and, for BBO, fluid energy as viscous stress (Basset history term, blue). (d) The time-derivative of the net work (done by each term in the BBO equation) gives the instantaneous power. Beads are neutrally buoyant ($\rho_s = \rho$) and driven by an identical "long" pulse of constant force $F_0 = \frac{1}{10}F_c$ applied from $t = 0$ to $\tau_0 = 10\tau_B$ (dotted lines).

That the Basset history force dominates Stokes drag on timescales shorter than the Brownian relaxation time agrees with the conclusions reached by Kheifets et al. [11], where the history force is seen to be dominant on the timescale of thermal velocity fluctuations. Conversely, when a force is applied to a BBO bead gradually, dissipation of the input energy is shared more equally between the Stokes drag and history terms.

For completeness, we also present results for the case of a pulse of "moderate" abruptness, having a duration exactly equal to $\tau_B$ [Fig. 2]. In this case, the Basset history force does in fact appear to change sign between 1 and 1.5 $\tau_B$, albeit only slightly [Fig. 2(b,d)]. The shift is much less dramatic than for when the pulse is longer than a Brownian relaxation time (i.e., $10\tau_B$). We conclude that applying a longer duration pulse therefore has the effect of storing energy in the vorticity field of the imaginary fluid bath surrounding the particle. This energy, which is really the energy associated with the stress field of the fluid, can be thought of as latent bead kinetic energy, since some of it will be returned to the bead in the future (in the form of momentum). Our simulations suggest that a square pulse of duration $\sim 3\tau_B = \tau_\nu$ causes the Basset term to maximize the return of latent energy to the bead during the relaxation phase. Our simulations suggest that the value of the pulse duration where latent energy return becomes important is around $\tau_\nu = 3\tau_B$.

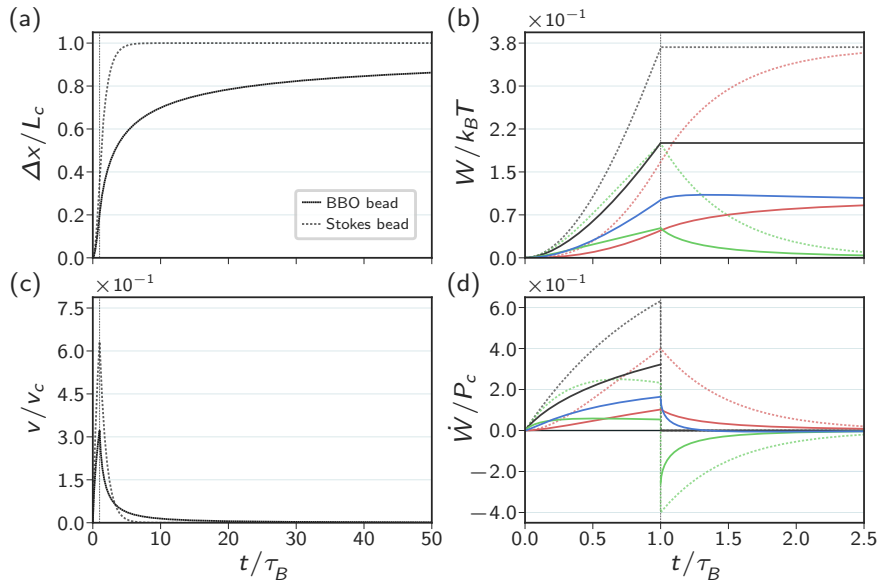

FIG. 2. Analytical solutions for BBO (solid lines) and Stokes (light dashed lines) beads for a single moderate pulse. (a) The net displacement $\Delta x(t)$ and (b) instantaneous velocity $v(t)$ both differ in the driving and relaxation phases. (c) The net work due to external driving (black) balances the kinetic energy (green), dissipated heat (Stokes term, orange) and, for BBO, fluid energy as viscous stress (Basset history term, blue). (d) The time-derivative of the net work (done by each term in the BBO equation) gives the instantaneous power. Beads are neutrally buoyant ($\rho_s = \rho_f$) and driven by an identical "long" pulse of constant force $F_0 = \frac{1}{10} F_c$ applied from $t = 0$ to $\tau_0 = 10\tau_B$ (dotted lines).

## SQUARE-WAVE DRIVING: VARYING DRIVING PERIOD AND PULSE DURATION

Using a longer-period forcing frequency, we see the same overall behavior as in Fig. 2 of the main text, where BBO beads exhibit a lower friction under time-periodic driving conditions [Fig. 3]. Again, the difference between the Stokes and BBO beads' effective friction is clearly established after only several forcing cycles.

Decreasing the forcing period to exactly one kinematic time ($\tau = 3\tau_B = \tau_\nu$), however, qualitatively changes the behavior [Fig. 4]. When the force is always on (i.e., pulse duration is equal to the period, $\tau_0 = \tau = \tau_\nu$), the effective friction for Stokes beads rapidly approaches a minimum value of 1 [dotted black line, Fig. 4(d)]. BBO beads, on the other hand, slowly approach this value, which is indicative of the algebraic decay of the memory kernel [solid black line, Fig. 4(d)]; the effective friction is about 1.5 times the value for a Stokes bead after several tens of Brownian relxation times, around 1.25 times the value after around $100\tau_B$, and within 10% at $500\tau_B$. However, more interesting behavior results when pulses are applied for a brief portion of a driving period. In particular, while BBO beads initially have a larger effective friction than Stokes beads, there are crossover points around roughly $50\tau_B$ and $150\tau_B$; after $500\tau_B$, the effective friction is about 15% lower for short pulses [$\tau_0 = \frac{1}{10}\tau$, blue, Fig. 4(d)] and 10% lower with very short pulses [$\tau_0 = \frac{1}{100}\tau$, red, Fig. 4(d)].

For shorter-duration pulses, the situation is more interesting. When the pulse duration is $\tau_0 = \frac{1}{10}\tau = \frac{1}{10}\tau$, the effective friction for BBO beads starts off above that of Stokes beads [blue lines, Fig. 4(d)]. However, after around $60\tau_B$, the BBO beads actually have a *lower* effective friction than Stokes beads. This value continues decreasing up to and beyond the duration of the simulation (i.e., $500\tau_B$), at which point the effective friction is around 30% lower for BBO beads than Stokes beads; Stokes beads attain their minimum effective friction after several tens of $\tau_B$. The situation is similar in the case of very short pulses, $\tau_0 = \frac{1}{100}\tau = \frac{1}{100}\tau$, where Stokes beads have a lower effective friction than BBO beads up to around $100\tau_B \simeq 30\tau_\nu$, after which the effective friction of BBO beads drops below that of Stokes beads [red lines, Fig. 4(d)]. After $500\tau_B$, BBO beads have about a 15% lower effective friction than Stokes beads. When the period is equal to the kinematic time, hydrodynamic memory still appears to produce a measurable advantage that manifest in a lower effective friction for BBO beads versus Stokes beads.

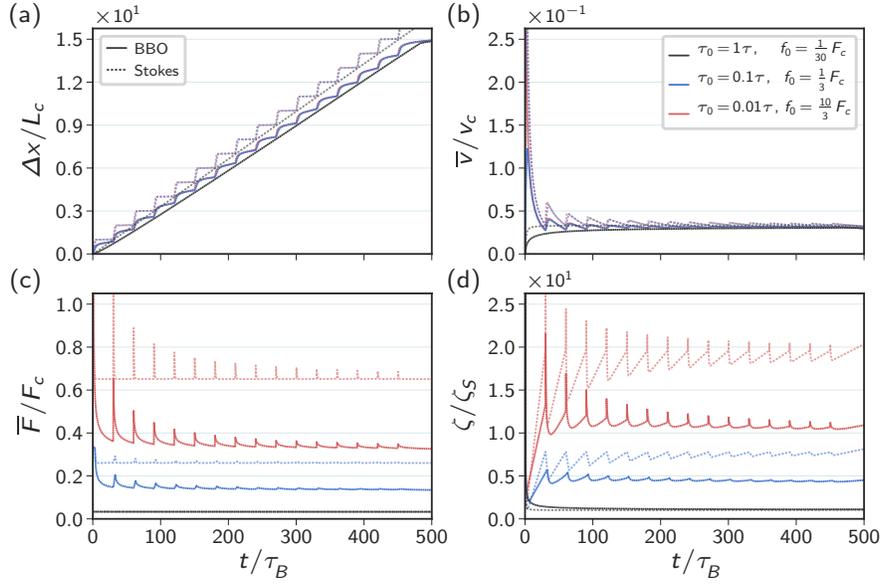

FIG. 3. Time-periodic forcing with a period ten times the kinematic time, $\tau = 10\tau_\nu$. Results are shown for BBO (solid lines) and Stokes (dotted) beads, and three values of the pulse duration: equal to the period (always-on) $\tau_0 = \tau$ (black), short pulses $\tau_0 = \frac{1}{10}\tau$ (blue), and very short pulses $\tau_0 = \frac{1}{100}\tau$ (red). Results are analogous to Fig. 2 in the main text: (a) displacement, (b) effective velocity, (c) effective driving force, and (d) effective friction. The impulse per cycle is held constant: $J_{\mathrm{cyc}} = f_0 \tau_0 = \mathrm{constant}$.

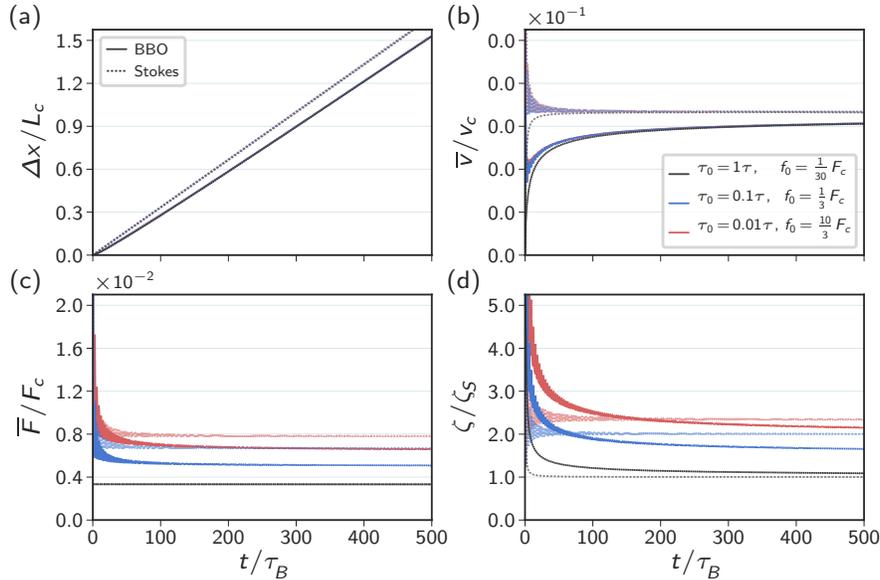

FIG. 4. Time-periodic forcing with period equal to the kinematic time, $\tau = \tau_\nu$. Results are shown for BBO (solid lines) and Stokes (dotted) beads, and three values of the pulse duration: equal to the period (always-on) $\tau_0 = \tau$ (black), short pulses $\tau_0 = \frac{1}{10}\tau$ (blue), and very short pulses $\tau_0 = \frac{1}{100}\tau$ (red). Results are analogous to Fig. 2 in the main text: (a) displacement, (b) effective velocity, (c) effective driving force, and (d) effective friction. The impulse per cycle is held constant: $J_{\mathrm{cyc}} = f_0 \tau_0 = \mathrm{constant}$.

Once the forcing period is decreased well below a kinematic time, $\tau = \frac{1}{10}\tau_\nu = 3\tau_B$ [Fig. 5], the differences between BBO and Stokes beads are much more muted. In particular, the effective friction for all beads is seen to settle to $\zeta \approx 1$ [Fig. 5(d)] rather quickly. The friction of BBO beads versus Stokes beads is only substantially larger for the first 100–200$\tau_B$, after which values are with 30% of the minimum friction value achieved by Stokes beads in the always-on case [i.e., when $\tau_0 = \tau$, dotted black lines, Fig. 5(d)].

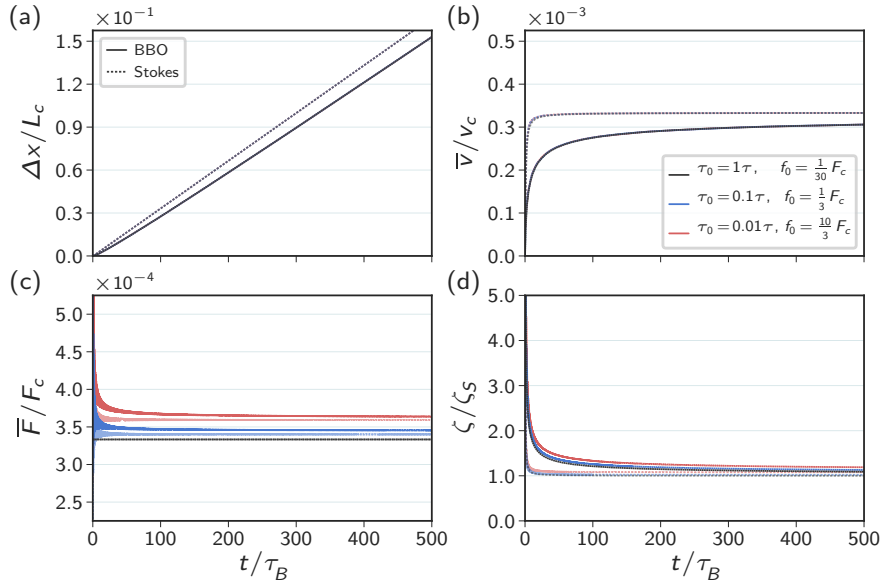

FIG. 5. Time-periodic forcing with period a tenth of the kinematic time, $\tau = \frac{1}{10}\tau_\nu$. Results are shown for BBO (solid lines) and Stokes (dotted) beads, and three values of the pulse duration: equal to the period (always-on) $\tau_0 = \tau$ (black), short pulses $\tau_0 = \frac{1}{10}\tau$ (blue), and very short pulses $\tau_0 = \frac{1}{100}\tau$ (red). Results are analogous to Fig. 2 in the main text: (a) displacement, (b) effective velocity, (c) effective driving force, and (d) effective friction. The impulse per cycle is held constant: $J_{\text{cyc}} = f_0 \tau_0 = \text{constant}$.

The reason for the differences in the high-frequency case can be rationalized by directly comparing the effective driving forces and velocities. Whereas the effective driving forces are comparable between BBO and Stokes beads subjected to identical forcing protocols in time [Fig. 5(c); values are within roughly 20% of one another after only $\sim 10\tau_B$], BBO beads have a measurably lower effective velocity than Stokes beads [Fig. 5(b); values differ by over 40% around $10\tau_B$ and are almost within 20% after $50\tau_B$]. However, if the BBO beads driven under this forcing protocol for a longer time (i.e., more cycles), it is expected that the effective friction achieved in steady state will eventually be the same as that achieved by Stokes beads. Overall, the differences between BBO and Stokes beads are nevertheless small as compared to the differences that arise for driving forces with a much longer period [as in Fig. 2 in the main text and Fig. 3 above]. Therefore, if Stokes beads have any transport advantage at all (by not having the Basset term present), it is for the scenario in which a high frequency force with period $\tau \lessapprox \tau_\nu$ is applied for a total length of time somewhat shorter than $100\tau_B$—before the effective friction of a BBO bead has had time to relax. Indeed, in the long-time limit, the effective friction for BBO beads would converge to the value achieved by Stokes beads. However, the overall differences between BBO and Stokes beads for high frequency forcing is relatively small as compared to when driving frequencies are below $\tau_\nu^{-1}$; at high frequencies, the effective friction is less sensitive to the pulse duration (i.e., abruptness) and is relatively close to the theoretical minimum value for always-on forcing.

To recapitulate, when the forcing frequency is below $\tau_\nu^{-1}$, then there is a measurable drop in the effective friction when memory is included. When the frequency is right at $\tau_\nu^{-1}$, there is a crossover after around 100 Brownian relxation times, where the BBO bead effective friction drops below that of the Stokes bead (for a given forcing function). In the case where the frequency is ten times *above* $\tau_\nu^{-1}$, the values for the friction all appear to approach a long-time value of $\zeta = \zeta_S$, with the effective friction of the BBO beads taking slightly longer than that for the Stokes beads, though the differences are relatively minor. It therefore seems that, for non-smoothly-varying driving forces, BBO beads are somewhat more difficult to transport for high-frequency driving (i.e., when BBO beads do not have enough time to reach a relatively constant steady-state velocity), but substantially easier to transport for low-frequency driving.

## TILTED SPATIAL POTENTIAL WITH GAPS: VARYING WAVELENGTH AND STEEPNESS

In order to test the robustness of our results, we performed long simulations $t_f = 2500\tau_B$ for both BBO and Stokes beads in the tilted spatial potential across five different values of the wavelength $\lambda$ and five values for the overall tilt (i.e., overall slope) $F_0$ of the tilted potential, for a total of 25 wavelength-tilt combinations. Then, for each pair of values $\lambda$ and $F_0$, we then tested the sensitivity of bead transport to the gap width by varying $d$ from $\frac{1}{20}\lambda$ to $\lambda$ in increments of $\frac{1}{20}\lambda$, simulating one bead for each value of the gap width. In total, we performed $25 \times 20 = 500$ simulations for each type of bead, with 1000 simulations in total.

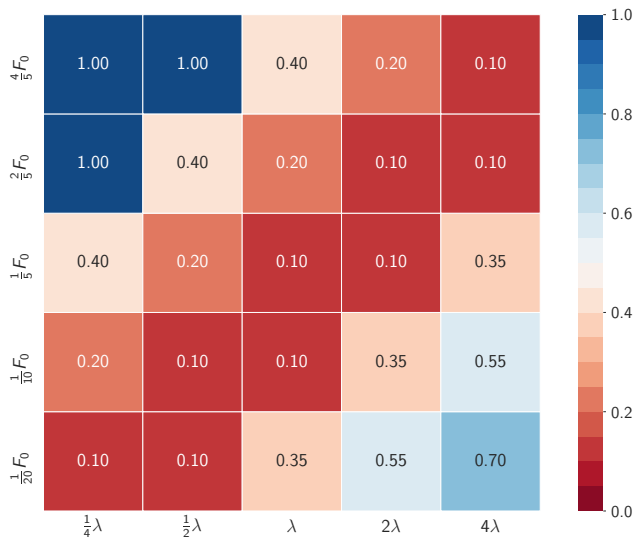

FIG. 6. Fraction of itinerant Stokes beads in the tilted spatial potential as a function of gap size for varying combinations of potential wavelength $\lambda$ and overall potential slope $F_0$. Equivalently, the heatmap value corresponds to the minimum gap fraction (i.e., the gap distance $d$ as the fraction of the wavelength) at which beads are localized in the gap. For instance, a value of 0.2 means that beads become stuck when the gap $d = 0.2\lambda$ (but achieved transport when $d = 0.15\lambda$ and smaller.)

Figure 6 and Figure 7 summarize the results for Stokes and BBO beads, respectively. A value of 1 means that, for the given combination of wavelength and tilt, 100% of the beads were able to clear the gaps and become itinerant. Equivalently, this means that the maximum gap width that was cleared by a bead was 95% of the wavelength of the potential. Similarly, a value of 0.1, for instance, means that only 10% of the beads (2 out of 20) acheived sustained transport; one of the beads encountered no gap (constant slope throughout the potential) and one of the beads cleared a gap that was 5% of the wavelength, i.e., a gap width $d = \frac{1}{20}\lambda$.

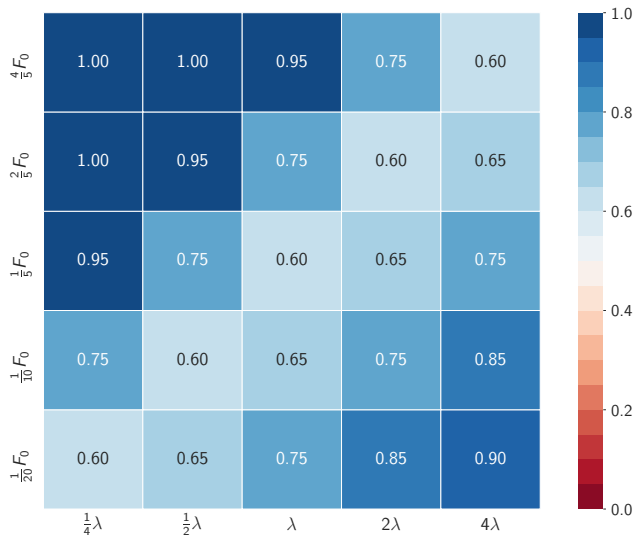

FIG. 7. Fraction of itinerant BBO beads in the tilted spatial potential as a function of gap size for varying combinations of potential wavelength $\lambda$ and overall potential slope $F_0$. Equivalently, the heatmap value corresponds to the minimum gap fraction (i.e., the gap distance $d$ as the fraction of the wavelength) at which beads are localized in the gap. For instance, a value of 0.2 means that beads become stuck when the gap $d = 0.2\lambda$ (but achieved transport when $d = 0.15\lambda$ and smaller.)

Since the gap width is set as a fraction of the wavelength, as the wavelength of the tilted potential is increased, the gap size $d$ increases proportionally. As a result, beads must clear a larger gap for a given gap fraction, so a larger amplitude is required to generate itinerant behavior for a given gap size. For the smallest wavelengths and steepest slopes, all beads, whether BBO beads or Stokes beads, were able to escape [top left, Fig. 6 and Fig. 7]. As the wavelength is increased, holding the potential slope constant, Stokes beads become localized relatively easily. For instance, when the potential slope is large $\frac{4}{5}F_0$ and the wavelength is $\lambda$, Stokes beads can only cross the gap when the gap width is less than 40% of a full wavelength, whereas BBO beads can cross a gap that is 90% of the wavelength; when the wavelength is $2\lambda$, the maximum crossable gap width for Stokes beads is less than 20% of the wavelength, but 75% for BBO beads [top row, Fig. 6 and Fig. 7].